\documentclass[11pt,b5paper]{article}
\usepackage{amssymb,amsfonts}
\usepackage{amsmath,latexsym}
\begin {document}
\centerline{\bf Two approaches for Helmholtz equation:
generalized} \centerline {\bf Darboux Transformation and the
method of ${\bar
\partial}$-problem}
\vskip0.5cm
\centerline{\bf E.Sh.Gutshabash}
\vskip0.3cm \centerline {Institute of Physics, St. -Petersburg
State University, Russia} \centerline{e-mail:
gutshab@EG2097.spb.edu} \vskip0.3cm

\begin{abstract}
Two approaches to solution of the two-dimensional Helmholtz
equation with a "wave
 number" are proposed. The results can be applied both in numerical
 areas of physics and in the
 theory of nonlinear equations.
 The first approach is based on the requirement of the
 covariance of equation under the generalized Darboux transformation
 (Moutard transformation). It allows to construct a new
 solution of equation, using a
 given initial solution of the equation. Simultaneously we
 obtain the "dressing" relation for the "wave number". The simplest
 examples of the approach are considered in detail.
 In the second approach the Green-Cauchy formula
 (the $\bar \partial$-method) is applied to reduce the solution of the
 equation to the solution of a system of singular integral
 equations.
\end{abstract}

\vskip0.5cm

\centerline{\bf {1. INTRODUCTION}} \vskip0.3cm

An intensive development of the methods in the theory of nonlinear
integrable equations stimulated as well the development of the so-called
dressing methods for the solution of a series of linear equations
having the important physical applications. One of the most
popular and effective methods is the method of Darboux
Transformation (DT) \cite{1}, which gives the exact
solutions of different one-dimensional equations. The situation for
the multidimensional equations is more complicate. In \cite{2}
the multidimensional variant of DT, which
depends on spatial derivatives, was considered, but such approach is
rather complicate and cumbersome.

In paper \cite{3} the method of integration of the non-stationary
Schrodinger and Fokker-Plank equations was proposed with using of
the Darboux-like anzats and introducing of some
functional-differential equation. It appeared, that in some sense
this approach is similar to the so-called Moutard transformation
(see, for example, \cite{4}) for the variable real "wave number"
(potential). In this case the functional-differential equation of
the second order is replaced by the system of two differential
equations of the first order for functional, included in Darboux's
anzats. For the complex "wave number" the situation is more
complicate - the solution can be found in some simple cases only.

In Section 2 of this paper the idea of a Darboux-type anzats -
Moutard transformation - is introduced in the example of the
two-dimensional Helmholtz equation. The main advantage of such
approach is independence of the procedure on the spatial
derivatives, and hence it can be straightforwardly extended onto
higher dimensionality of space.

In Section 3 the simplest classical problem of diffraction of the
electromagnetic waves on a half-plane is considered. We
reformulate this problem in terms of a so-called method of ${\bar
\partial}$-problem. The idea of such approach (in another
formulation of the problem) belongs to V.D.Lipovskii \cite{5}. The
powerful and effective possibilities of this method for
integration of multidimensional systems such as
Kadomtsev-Petviashvili-2 equation, Devey-Stewartson-2 system and
so on were demonstrated (see, for example, \cite{6}-\cite{7}).
Apparantely, up to now this approach was not applied in the theory
of diffraction. Similar to the Wiener-Hopf method, it allows to
formulate the initial boundary problem in terms of the integral
equations, which admit the explicit exact solutions, and
simultaneously to obtain a series of useful and important
relations for this problem.

\vskip0.4cm \centerline {\bf 2. The generalized Darboux
Transformation} \vskip0.4cm a). \enskip The two-dimensional
Helmholtz's equation belongs to a wide class of the equations of
the following form (see, for example, \cite{8}):
\begin{equation} \label{1}
\Psi _ {pq} = V (p, q) \Psi,
\end{equation}
where $ V = V (p, q) $ is the given, in general case,
complex-valued function. Here and below we use the notations: $
\Psi_q = \partial \Psi/\partial q, \ldots .$

In the simplest particular case: $ p = x, \:q = y, \:V (x,
y) \equiv 1 $, this equation has the general solution
\begin{equation}\label{2} \Psi (x, y) = \int _ 0 ^ xf _ 1 (s) J _ 0 (2i\sqrt {y (x-s)}
\:)\:ds + \int _ 0 ^ yf _ 2 (s) J _ 0 (2i\sqrt {x (y-s)} \:) ds +
$$ $$+[f _ 1 (0) + f _ 2 (0)] J _ 0 (2i\sqrt {xy} \:),
\end{equation}
where $ J _ 0 (.) $ is the Bessel's function, and $ f_1, \:f_2 $
are some differentiable functions.

 The class of such equations is important both for
the differential geometry and for a wide variety of the physical
problems. Indeed, for example, for $ p = x + t, \:q = x-t $ we
have the heterogeneous wave equation:
\begin{equation}\label{3}
\Psi _ {tt} -\Psi _ {xx} = V (x, t) \Psi,
\end{equation}
and for $ p = z = x + iy, \:q = \bar p $ we obtain
the Helmholtz equation itself:

\begin{equation}\label{4}
 \triangle \Psi = V (x, y) \Psi. \end{equation}
 We shall use it also in the complex form ($ \partial _ z = (1/2) (\partial _
x-i\partial _ y), \: {\partial _ {\bar z}} = (1/2) (\partial _ x +
i\partial _ y)$)

\begin{equation}\label{5}
 \Psi _ {z \bar z} = \frac {1} {4} V (z, \bar z) \Psi. \end{equation}

The equation (\ref{4}{(5)}) arises in the Quantum Mechanics
(two-dimensional stationary Schrodinger equation), theory of
diffraction of electromagnetic waves, an acoustic diffraction and
other problems. From the nonlinear equations point of view the
interest to (\ref{4}) is caused by his role as the associated
linear system for the well-known nonlinear completely integrable
Novikov-Veselov equation \cite{9}.

It is not difficult to check by direct calculation that if $ \Phi
= \Phi (z, \:\bar z), \:\chi = \chi (z, \:\bar z) $ are two
solutions of equation (\ref{4}), then two important relations are
true:

\begin{equation}\label{6}
 ( \Phi _ z \chi) _ {\bar z} = (\Phi _ {\bar z} \chi) _ z, \:\:\:
(\Phi _ z\chi-\Phi \chi _ z) _ {\bar z} = (\Phi \chi _ {\bar z}
-\Phi _ {\bar z} \chi) _ z.  \end{equation} It follows from the
second of them, that the integral

\begin{equation}\label{7}
 \omega (\Phi, \:\chi) = \int _ {(z _ 0, \: {\bar z} _ 0)} ^ {(z,
\:\bar z)} (\Phi _ {z} \chi-\chi _ {z} \Phi) dz + (\chi _ {\bar z}
\Phi-\chi \Phi _ {\bar z}) d {\bar z},  \end{equation}
does not depend on the integration path.

 b). \enskip We shall proceed to the analysis of the equation
 (\ref{5})
 from the generalized Darboux Transformation point of view.
 Let $ \Psi, \:\Psi _ 1 $ are two linearly independent solutions of (\ref{5}).
 Following \cite{1,3}, we assume that

\begin{equation}\label{8}
 { \tilde \Psi} (z, {\bar z}) = \frac {\Omega (\Psi, \Psi _ 1)}
{\Psi _ 1}, \end{equation} where $ \Omega = \Omega (\Psi, \Psi _
1) = \Omega (z, \bar z) $ is the functional, given on direct
product of two copies of spaces  of "wave functions ". After
substitution of the anzats (\ref{8}) in equation (\ref{5}), and
requiring the covariance of this equation under transformation $
\Psi \to \tilde \Psi, \:U \to \tilde U $, we obtain the equation,
which do not contain already the potential:

\begin{equation}\label{9}
 \Omega _ {z {\bar z}} - (\ln \Psi _ 1) _ {\bar z} \Omega _ z -
(\ln \Psi _ 1) _ z\Omega _ {\bar z} + q _ 1\Omega = 0.
\end{equation}
Here $ q _ 1 = q _ 1 (z, \bar z) $ is a complex-valued function (
"parameter" of separation of variables), introduced in order to
obtain the real potential $ {\tilde V} (z, \bar z) $ for the case
of $ V (z, \bar z) = {\bar V} (z, \bar z) $ (conservative medium).
If $ V (z, \bar z) \ne {\bar V} (z, \bar z) $ and $ {\tilde V} (z,
\bar z) \ne {\bar {\tilde V}} (z, \bar z) $ (nonconservative
medium), one may take  $q_1=0$ in (\ref{9}).

Equation (\ref{9}) allows the physical interpretation: it looks
like the two-dimensional stationary Schrodinger equation for
charged particle in a non-homogenious stationary electromagnetic
field { \footnote {Author is grateful to I.V.Komarov for this
remark.}}. Actually, this equation can be written as (see, for
example, \cite{10}):

\begin{equation}\label{10}
\triangle \psi-2i ({\bf A} \nabla \psi) - ({\bf A} ^ 2 + \phi-E)
\psi = 0, \end{equation}
where $ {\bf A} (x, \:y) = (A _ 1, \:A _
2) $ is the vector potential, with the
gauge condition $ {\mathrm div} \: {\bf A} = 0, \:\phi = \phi (x, \:y)$ is
the scalar potential, $E$ is the energy value, and the
system of units $ \hbar = c = e = 1, \:m = 1/2 $ was used
($m$ is the mass of the particle). In variables $ z, \: {\bar z} :$

\begin{equation}\label{11}
 \psi _ {z {\bar z}} -\frac {1} {4} (iA _ 1-A _ 2) \psi _ z-\frac
{1} {4} (iA _ 1 + A _ 2) \psi _ {\bar z} -\frac {1} {4} [{\bf A} ^
2 + \phi-E] \psi = 0. \end{equation} Comparing (\ref{11}) and
(\ref{9}), we see, that analogues of coefficients at $ \Omega _ z,
\:\Omega _ {\bar z} $ are expressed as linear combinations of the
components of the vector-potential, and the analogue of function
$q_1$ - in terms of the square of its module and of the scalar
potential.

Besides, the separation of variables leads to the
following dressing relation for potential (wave number):

\begin{equation}\label{12}
 \tilde V (z, \bar z) = V (z, \bar z) -8 (\ln \Psi _ 1) _ {z\bar
z} -4q _ 1 (z, \bar z). \end{equation}
 In the conservative case the requirement of reality reads:

\begin{equation}\label{13}
 ( {\mathrm arg} \:\Psi _ 1) _ {z \bar z} = -4\: {\mathrm Im} \:q
_ 1. \end{equation}
 For the double dressing procedure, with

\begin{equation}\label{14}
 { \tilde {\tilde \Psi}} (z, \:\bar z) = \frac {\Omega (\Psi _ 2
[1], \:\Psi [1])} {\Psi _ 2 [1]}, \end{equation}
 and taking into account (\ref{12}), we obtain

\begin{equation}\label{15}
 { \tilde {\tilde V}} (z, \:\bar z) = V-8 (\ln \Omega (\Psi _ 1,
\:\Psi _ 2))_{z \:\bar z} -4 \{q _ 1 + q _ 2 [1] \},
\end{equation}
where $ \Psi _ 2 [1] $ is the fixed solution of equation (\ref{5})
with $ \Psi \to \Psi [1], \:V \to V [1] $, and $ \Psi _ 2 [1] =
\Omega (\Psi _ 1, \:\Psi _ 2) /\Psi _ 1, \:\Psi _ 2 $ is some
fixed solution of equation (\ref{5}).

It is clear, that similar to the case of standard DT, the dressing
procedure proposed here can be multiply iterated. Setting $ \Psi =
\Psi [0], \:\tilde \Psi = \Psi [1], \ldots, V = V [0], \:\tilde V
= V [1], \ldots $, we have the chains: $ \Psi [0] \to \Psi [1] \to
\ldots \to \Psi [N] \to \ldots $, and $ V [0] \to V [1] \to \ldots
\to V [N] \to \ldots $. In particular, for the $ N $-fold
iteration of the dressing procedure, one will obtain the
relations, expressed in terms of the functions $ \Omega (\Psi _ i,
\:\Psi _ j) $ only \cite{1}.

c). \enskip Now some simplest examples of
realization of the proposed approach will be considered.

1. Let us take $ V (z, \bar z) = 0 $ in (\ref{5}). Then $ \Psi _ j
(z, \:\bar z) = \Psi _ j ^ {(1)} (z) + \Psi _ j ^ {(2)} (\bar z),
\: \Psi _ j ^ {(1)}, \:\Psi _ j ^ {(2)} $ are abitrary functions,
$j=1, \:2, \ldots, N, \ldots $. The relation (\ref{12}) takes a
form:

\begin{equation}\label{16}
 V [1] = -8 [\ln \left.\left (\Psi _ 1 ^ {(1)} (z) + \Psi _ 1 ^
{(2)} (\bar z) \right.\right)] _ {z \bar z} -4q _ 1.
\end{equation}
In the case $V [1]={\bar V [1]}$ the following condition

\begin{equation}\label{17}
 { \mathrm Im} \:q _ 1 = i\:\left [\frac {\Psi _ z ^ {(1)} \Psi _
{\bar z}^{(2)}} {(\Psi^{(1)} + \Psi^{(2)})^2} -\frac {{\bar \Psi}
_ {\bar z} ^ {(1)} {\bar \Psi} _ z ^ {(2)}} {({\bar \Psi} ^ {(1)}
+ {\bar \Psi}^{(2)})^2} \right] \end{equation}
must be satisfied.
On the next step we have:

\begin{equation}\label{18}
 V [2] = V-8 (\ln \Omega (\Psi _ 1, \:\Psi _ 2)) _ {z \bar z} -4
[q _ 1 + q _ 2 [1]), \end{equation} and the functional $ \Omega
(\Psi _ 1, \:\Psi _ 2) $ satisfies the equation (\ref{9}), if to
replace $ \Psi \to \Psi_1,\:\Psi_1 \to \Psi_2,\:\:q_1 \to q _ 2
[1] $ in it. It is not difficult to check, that for $q_1 = q_2
[1]=0$ this equation is solved by the function $ \Omega (\Psi _ 1,
\:\Psi _ 2) = \omega (\Psi _ 1, \:\Psi _ 2) $, where the
expression for $ \omega (\Psi _ 1, \:\Psi _ 2) $ is defined in
(\ref{7}). After the simple calculations we obtain:

\begin{equation}\label{19}
 \Omega (\Psi _ 1, \:\Psi _ 2) = \gamma _ 0 + 2 (\Psi _ 1 ^ {(1)}
\Psi _ 2 ^ {(2)} -\Psi _ 2 ^ {(1)} \Psi _ 1 ^ {(2)}) +
$$
$$
+ \int _ {(z _ 0, \: {\bar z} _ 0)} ^ {(z, \:\bar z)} (\Psi _ {1z}
^ {(1)} \Psi _ 2 ^ {(1)} -\Psi _ {2z} ^ {(1)} \Psi _ 1 ^ {(1)}) dz
+ (\Psi _ {2\bar z} ^ {(2)} \Psi _ 1 ^ {(2)} -\Psi _ {1\bar z} ^
{(2)} \Psi _ 2 ^ {(2)}) d {\bar z}, \end{equation} where $ \gamma
_ 0 $ is a complex constant. This expression, without using of the
equation (\ref{9}), was obtained for the first time in \cite{1}.

2. Let's assume in (\ref{4}) that $ V = V (x, y) = 1 $ at $
-\infty < x < + \infty, \: y> 0 $. Applying the Fourier
transformation in the variable $x$, we shall construct the
solution $ \Psi _ j\: (j = 1,2, \ldots, N, \ldots), $ which is
bounded in $x$ and goes to zero at $y \to \infty$:

\begin{equation}\label{20}
 \Psi _ j (x, \:y) = \frac {1} {2\pi} \int _ {-\infty} ^ {+
\infty} dk \: e^{ikx-\sqrt {k^2+1} y} A _ j (k).
\end{equation}
Here $A_j(k)$ are some functions (functional parameters), such
that $ A_j (k) \in {\mathbb L}_2 (-\infty, \infty)$. In the real
case one obtains from (\ref{19}) $ A _ j (k) = {\bar A} _ j (-k) $
(in case of complex $ k $ we assume, that $ {\mathrm Re} \: {\sqrt
{k ^ 2 + 1}} > 0 $). Then

\begin{equation}\label{21}
 V [1] = 1-2\triangle \ln \{\int_{-\infty}^{+ \infty} dk \:
e ^{ikx-\sqrt {k ^ 2 + 1} y} A _ 1 (k) \}. \end{equation}
 The analogue of the second relation in (\ref{6}) yields:

\begin{equation}\label{22}
( \Psi _ x\Psi _ 1-\Psi _ {1x} \Psi _ x) _ x = (\Psi _ {1y}
\Psi-\Psi _ 1\Psi _ y) _ y.
\end{equation}
Therefore the solution $ \Psi [1] $ is determined by a relation
(\ref{8}) at $ \Omega (\Psi, \:\Psi _ 1) = \omega (\Psi, \:\Psi _
1) $, where

\begin{equation}\label{23}
 \omega (\Psi, \Psi _ 1) = \int _ {(x _ 0, \:y _ 0)} ^ {(x, y)}
(\Psi _ {1y} \Psi-\Psi _ 1\Psi _ y) dx + (\Psi _ x\Psi _ 1- \Psi _
{1x} \Psi) dy.
\end{equation}
On the next step we obtain:

\begin{equation}\label{24}
 V [2] = 1-2\triangle \: (\ln \Omega (\Psi _ 1, \:\Psi _ 2)) .
\end{equation}
In the particular case of $ A _ j (k) = c _ j \delta (k-p _ j) + d
_ j \delta (k-r _ j), $ with $ \delta (.) -$ the Dirac  delta-
function, $c_j, \:d _ j, \:p _ j, \:r _ j \in {\mathbb R} -$
constants,\:$j = 1, 2, \ldots, N, \ldots $, taking (due to
linearity of the equation (\ref{5})) the real part of the solution
$ \Psi _ j $ we shall have:

\begin{equation}\label{25}
 \Psi _ j (x, \:y) = c _ je ^ {-\sqrt {p _ j ^ 2 + 1} y} \cos p _
jx + d _ je ^ {-\sqrt {r _ j ^ 2 + 1} y} \cos r _ jx.
\end{equation}
Thus, the dressing formula (\ref{15}) gives:

\begin{equation}\label{26}
 V [1] = 1-2\triangle \ln\:(c_1e^{-\sqrt {p_1^2+1} y}
\cos p_1x + d_1e^{-\sqrt {r_1^2+1} y} \cos r_1x) .
\end{equation}
Completely similarly, it is possible to obtain from (\ref{15}),
(\ref{24}) the explicit representation for $ V[2]$, though rather
cumbersome.

3. Let us consider a case, when the initial solution is determined
by the Coulomb potential, i.e. in (\ref{4}) we assume that $ V (x,
\:y) = ({\sqrt {x^2+y^2}} \:)^{-1}, \:x \in (-\infty, \:\infty),
\:y \in (-\infty, \:\infty) $. Solving this equation in the polar
coordinates and coming back to the cartesian ones, (with
requirement that the solution is finite at $ x = y = 0 )$, we
obtain \cite{11}:

\begin{equation}\label{27}
 \Psi _ m (x, y) \sim e ^ {im \arctan \frac {y} {x}} J _ {2i
{\sqrt {m}}} \left ( 2i (x ^ 2 + y ^ 2) ^ {\frac {1} {4}} \right),
\: \:\:m = 1, \:2, \ldots, \:N, \ldots,
\end{equation}
where $ J _ {\nu} (.) $ is the Bessel function. Then for
real potential we find:

\begin{equation}\label{28}
 V [1] = \frac {1} {\sqrt {x ^ 2 + y ^ 2}} -2\triangle \ln \left
\{{\mathrm Re} \left [\:e ^{im \arctan \frac {y} {x}} J _ {2i
{\sqrt {m}}} \left ( 2i (x ^ 2 + y ^ 2) ^ {\frac {1} {4}} \right)
\right] \right \}.
\end{equation}
It is clear now, that already on a first step we obtain
the potential which is not amenable for separation of variables.
Hence, all subsequent
functions of the chain $ \Psi [k], \:V [k] $ obey this property as well.

Potentials of the form (\ref{26}) and (\ref{28}), and of the
corresponding chains, are integrable, i.e. the equations, produced
by them, have the exact solutions.

In two following examples we consider a case $q_1 \neq 0$ when the
equation (\ref{9}) allows the simplest solutions.

4. Let $ \Omega _ z = 0 $, i.e. $ \Omega $ is an antiholomorphic
function in some domain $ {\mathbb D} \subset {\mathbb C} $. From
(\ref{9}) we have:

\begin{equation}\label{29}
 ( \ln \Omega) _ {\bar z} = \frac {q _ 1} {(\ln \Psi _ 1) _ z}
\:\:\: \mbox {at the condition, that} \:\:\: (\frac {q _ 1} {(\ln
\Psi _ 1) _ z}) _ z = 0.
\end{equation}

To solve this equation we use the formula of $ \bar
\partial $-problem (see, for example, \cite{6}). Assuming that

\begin{equation}\label{30}
 \Omega _ {| \partial {\mathbb D}} = C _ 0,
\end{equation}
where $ C_0 $ is some complex constant, we find:

\begin{equation}\label{31}
 \ln \Omega (\bar z) = C _ 0 + \frac {1} {2\pi i} \int \int _
{\mathbb D} \frac { q _ 1} {(\ln \Psi _ 1) _ {\zeta} (\zeta-z)}
\:d\zeta \wedge d {\bar \zeta}.
\end{equation}
Here $ \zeta = \zeta _ R + i\zeta _ I, \:d\zeta \wedge d {\bar
\zeta} = -2i d\zeta _ R\: d\zeta_I$. The correctness of this
expression follows from the well-known relation of the theory of
generalized functions ($ {\bar \partial} \equiv \partial _ {\bar
z}$):

\begin{equation}\label{32}
 { \bar \partial} (\frac {1} {\pi (z-z _ 0)}) = \delta (z-z _ 0).
\end{equation}

5. Analogously, let $ \Omega _ {\bar z} = 0 $, i.e. $ \Omega $ is
an holomorphic function in $ {\mathbb D} \subset {\mathbb C} $.
From (\ref{9}) we obtain:

\begin{equation}\label{33}
 ( \ln \Omega) _ z = \frac {q _ 1} {(\ln \Psi _ 1) _ {\bar z}}
\:\:\:\:\mbox {at the condition, that} \:\:\:\: (\frac {q _ 1}
{(\ln \Psi _ 1) _ {\bar z}}) _ {\bar z} = 0.
\end{equation}
Assuming, that

\begin{equation}\label{34}
 \Omega _ {| \partial {\mathbb D}} = C _ 1,
\end{equation}
where $ C_1$ is complex constant, from (\ref{33}) we have

\begin{equation}\label{35}
 \ln \Omega (z) = C _ 1 + \frac {1} {2\pi i} \int \int _ {\mathbb
D} \frac { q _ 1} {(\ln \Psi _ 1) _ {{\bar \zeta}} ({\bar \zeta} -
{\bar z})} \:d\zeta \wedge d {\bar \zeta},
\end{equation}

d). \enskip Now we shall find the quantum-mechanical sense of the
transformations (\ref{8}). For this purpose we shall consider more
general, than (\ref{5}), equation:

\begin{equation}\label{36}
\Psi _ {z \bar z} = \frac {1} {4} (V (z, \bar z) -\lambda) \Psi,
\end{equation}
where $\lambda \in \mathbb {C} $ is a complex parameter. It is
possible also to use here the transformation (\ref{8}) and
dressing relation (\ref{15}). Let us assume that $V (z, \bar z) =
z {\bar z} $, corresponding to the two-dimensional isotropic
harmonic oscillator. Then the equation (\ref{36}) can be rewritten
as

\begin{equation}\label{37}
 -4\Psi _ {z\:\bar z} (z, \:\bar z, \:\lambda) + z {\bar z} \Psi
(z, \:\bar z, \:\lambda) = \lambda \Psi (z, \:\bar z, \:\lambda).
\end{equation}
For $ \lambda = \lambda _ 1 = -2 $ this equation has the solution
$ \Psi _ 1 (z, \:\bar z) = \exp (\frac {1} {2} z {\bar z}) $, and
agrees with (\ref{12}) $ \tilde V (z, \:\bar z)=z{\bar z}-4-4q_1
$. Thus we obtain the equation

\begin{equation}\label{38}
 -4\Psi _ {z \bar z} [1] (z, \:\bar z, \:\lambda) + (z {\bar z}
-4q _ 1) \Psi [1] (z, \:\bar z, \:\lambda) = (\lambda + 4) \Psi
[1] (z, \:\bar z, \:\lambda).
\end{equation}
Comparing (\ref{37}) and (\ref{38}) for $q_1=0$, we find

\begin{equation}\label{39}
 \Psi [1] (z, \:\bar z, \:\lambda) = \Psi (z, \:\bar z, \:\lambda
+ 4),
\end{equation}
i.e. the transformation (\ref{8}), together with the requirement
of covariance, acts as the quantum-mechanical creation operator of
particles {\footnote {Accordingly for $ \lambda = \lambda _ 1 = +
2, \:\Psi _ 1 = \exp {(-\frac {1} {2} z {\bar z})} $, we have the
annihilation operator.}}, and the quantity $ -4q _ 1 $ can be
interpreted as the energy level shift (in a complex case). In this
sense we have two-dimensional generalization of the standard DT
\cite{1}.

Here the equation for functional $\Omega$ follows from (\ref{9}):

\begin{equation}\label{40}
 \Omega _ {xx} + \Omega _ {yy} -2x\:\Omega _ x-2y\:\Omega _ y + q
_ 1\Omega = 0.
\end{equation}
This equation, which can be called the generalized Helmholtz
equation, is known in the mathematical literature \cite{12}. It
arises under consideration of orthogonal polynomials in two
independent variables. For $q_1=2(n + m)$ it has the solution

\begin{equation}\label{41}
\Omega (x, \:y) = {\cal F}_{n + m, m} (x, \:y)=H_n (x) H_m (y),
\end{equation}
where $ H_l (.) $ are polynomials of Chebychev-Ermit, $m, \:n = 0,
\:1, \ldots $. Here the generalized condition of orthogonality is

\begin{equation}\label{42}
 \int _ {-\infty} ^ {\infty} \int _ {-\infty} ^ {\infty} h (x,
\:y) {\cal F} _ {nm} (x, \:y) {\cal F} _ {ks} (x, \:y) \:dx\: dy =
\delta _ {n-m, k-s} \:\delta _ {ms},
\end{equation}
where $h (x, \:y)=\exp {\{-(x^2+y^2) \}} $ is the weight function,
and $ \delta _ {ij} $ is the Kronecker symbol.

For $q_1=0$ the solution of equation (\ref{40}) also can be
factorized. Setting $ \Omega (x, \:y) =\Omega_1(x))\Omega_2(y) $,
we have:

\begin{equation}\label{43}
\Omega _ {1xx}-2x\Omega _ {1x}-\beta \Omega_1 = 0,
\:\:\:\:\:\:\Omega_{2yy}-2y\Omega_{2y}+\beta \Omega_2=0,
\end{equation}
where $ \beta $ is the "separation of variables parameter". For $
\beta = 2n $, where $ n = 1, \:2, \ldots ,$ the second of these
equations has the solution in terms of the Chebychev-Hermit
polynomials: \enskip $\Omega_2(y) = H _ n (y) $. Setting $ \Omega
_ 1(x) \equiv {\cal P} (x) = \sum _ {k = 0} ^ {\infty} a _ k x ^ k
$, from the first of the equations (\ref{43}) we find the
reccurent relation:

\begin{equation}\label{44}
a_{k + 2} = \frac {2 (k + n)} {(k + 2) (k + 1)} a _ k,
\end{equation}
where the numbers $ a_0, \: a_1$ remain arbitrary, and the power
series for  function $ {\cal P} (x) $ converges uniformly on the
entire axis.

Thus, in the dressing procedure the solution of equation
(\ref{36}) is

\begin{equation}\label{45}
 \tilde \Psi (x, \:y) = \Psi [1] (x, \:y) = c _ ne ^ {x ^ 2 + y ^
2} {\cal P}(x)H_n (y),
\end{equation}
where $c_n$ is a constant. Besides that, similarly to the
one-dimensional case (one-dimensional stationary Schrodinger
equation), the states described by the relation (\ref{45}) are not
normalizable. The detailed investigation of the creation and
annihilation operators in two and more spatial dimensions was
performed in the monograph \cite{10}.

\vskip0.8cm

\centerline {\bf 3.  A method of $\bar \partial $ - problem.}
 \vskip0.5cm

In this Section we consider the application
of the $ {\bar
\partial} $-problem method for the simplest case of diffraction on
half-planes (see, for example, \cite{14}):

\begin{equation}\label{46}
 \triangle u + k ^ 2u = 0,
\end{equation}
and {\footnote {Strictly speaking, it is necessary to add to
(\ref{47})-(\ref{48}) the condition on function $u$ on an edge.}}

\begin{equation} \label{47}
u = -u ^ {(i)} (x, \:y) = - e ^ {-ikx\cos \theta} \:\:\: \mbox
{at} \:\:\: x < 0, \:\:y \to \pm 0,
\end{equation}
\begin{equation} \label{48}
u = O (\frac {1} {\sqrt {r}}), \:\:\:\frac {\partial u} {\partial
r} -iku = o (\frac {1} {\sqrt {r}}) \:\: \mbox {at} \:\:\: r =
\sqrt {x ^ 2 + y ^ 2} \to \infty.
\end{equation}
Here $ u = u (x, \:y) = u ^ {(tot}) (x, \:y) -u ^ {(i)} (x, \:y),
\:u ^ {(tot)} $ is the total potential, $ u ^ {(i)}) $ is the
potential of a falling wave, $ u^ {(i)} = \exp {(-ikx\cos
\theta-iky\sin \theta)}, \:\triangle $ is the two-dimensional
Laplace operator, $ \:\: k = k _ 1 + ik _ 2, \:k _ {1}, \:k _
2> 0, \:0 \leq \theta \leq \pi $.

Let us introduce the variables $z$ and $ \bar z $, and let us
assume $ u (z, \:\bar z) = v(z\:\bar z) \: {\mathrm \exp} \: (k _
{-} z + k _ {+} {\bar z}) $, where $k _ {+} = - (i/2) k \:
{\mathrm \exp} \: (i\theta), \:k _ {-} = - (i/2) k \: { \mathrm
\exp} \: (-i\theta) $. Then in the $z$ - representation the
problem (\ref{46})-(\ref{48}) can be written in terms of function
$ v = v (z, \: {\bar z}) \equiv v (z, \:\bar z, \:k _ {+}, \:k _
{-}) $ as

\begin{equation}
\label{49} v _ {z {\bar z}} + k _ {+} v _ z + k _ {-} \:v _ {\bar
z} = 0,
\end{equation}

\begin{equation} \label{50}
 v (z, \: {\bar z}) = -1 \:\:\:\mbox {at} \:\:\:\: z + {\bar z} <
0, \:z - {\bar z} \to \pm i0.
\end{equation}
The radiation condition (\ref{48}) gives

\begin{equation} \label{51}
 v (z, \: {\bar z}) = o (1) \:\:\: \mbox {at} \:\: | z | \to
\infty.
\end{equation}

 It is not difficult to obtain the symmetry property from the equation (\ref{49}):

\begin{equation} \label{52}
v(z, \: {\bar z}; \:k _ {+} \:k _ {-}) = v ({\bar z}, \:z; \:k _
{-}, \:k _ {+}).
\end{equation}

In order to derive the integral equations we
note, that for any complex-valued function $ g (z, \bar z) \in C
(\bar {\mathbb D}) \bigcap C ^ 1 (\mathbb D), \: {\mathbb D}
\subset { \mathbb C}$ the Green formulas \cite{13} can be used:

\begin{equation} \label{53}
\oint _ {\partial \mathbb D} g d\zeta = -\int _ {\mathbb D} \int g
_ {\bar \zeta} \:d\zeta \wedge
 {\bar \zeta}, \:\:\:\oint _ {\partial \mathbb D} g d {\bar
\zeta} = \int _ {\mathbb D} \int g _ {\zeta} \:d\zeta \wedge d
{\bar \zeta}.
\end{equation}

Setting $ g (\zeta, \bar \zeta, \:l _ 1, \:l _ 2) = g _ 1 (\zeta,
\bar {\zeta}, \:l _ 1, \:l _ 2) = w (\zeta, \: {\bar \zeta}, \:l _
1, \:l _ 2) / (\zeta-z), \:\:g (\zeta, {\bar \zeta}, \:l _ 1, \:l
_ 2) = g _ 2 (\zeta, \bar \zeta, \:l _ 1, \:l _ 2) = w (\zeta, \:
{\bar \zeta}, \:l _ 1, \:l _ 2) / ({\bar \zeta} -\bar z), \:\:w
(\zeta, \: {\bar \zeta}) \equiv w (\zeta, \:\bar \zeta, \:l _ 1,
\:l _ 2)\in C (\bar {\mathbb D}) \bigcap C ^ 1 (\mathbb D)$, where
$ l _ 1, \:l _ 2 \in {\mathbb C} $ are parameters, applying to
these functions the Cauchy's formula with paths $
\partial { \mathbb D} \bigcup | \zeta-z | = \epsilon _ 1 $, and $
\partial {\mathbb D} \bigcup | {\bar \zeta} - {\bar z} | =
\epsilon _ 2 $ accordingly, where $ \epsilon _ 1, \:\epsilon _ 2
> 0 $, and passing to limits $ \lim \epsilon _ {1,2} = 0 $,
we obtain the integral representations:

\begin{equation} \label{54}
w (z, \: {\bar z}) = \int _ {\partial \mathbb D} \frac {d\zeta}
{2\pi i} \:\frac {w (\zeta, \: {\bar \zeta})} {\zeta-z} + \int _
{\mathbb D} \int \frac {d\zeta \wedge d {\bar \zeta}} {2\pi i}
\:\frac {\partial _ {\bar \zeta} w (\zeta, \: {\bar \zeta})}
{\zeta-z},
\end{equation}

\begin{equation} \label{55}
 w (z, \: {\bar z}) = -\int _ {\partial \mathbb D} \frac {d
{\bar \zeta}} {2\pi i} \:\frac {w (\zeta, \: {\bar \zeta})} {{\bar
\zeta} - {\bar z}} + \int _ {\mathbb D} \int \frac {d {\zeta}
\wedge d {\bar \zeta}} {2\pi i} \:\frac {\partial _ {\zeta} w
(\zeta, \: {\bar \zeta})} {{\bar \zeta} - {\bar z}}.
\end{equation}

To apply these representations, equation (\ref{49}) can be
rewritten as

\begin{equation} \label{56}
\partial _{\bar z} (\partial_z+k_{-}) v (z, \:\bar z)
= -k _ {+} \partial _ zv (z, \:\bar z), \:\:\:\partial _ z
(\partial _ {\bar z} + k _ {+}) v (z, \:\bar z) = -k _ {-}
\partial _ {\bar z} v (z, \:\bar z) .
\end{equation}
We assume, that

\begin{equation} \label{57}
 v_z (z, \: {\bar z}) = o (1), \:\:v _ {\bar z} (z, \: {\bar z})
= o (1) \:\:\mbox {at} \: | z | \to \infty.
\end{equation}
Choosing the path as $ \partial {\mathbb D} = \Gamma\bigcup C _
R,$ where $ \:C _ R $ is the circle of radius $ R $, and contour $
\Gamma $ consists of the straight line $ y = \pm i0 $, at $ x < 0
$, and the half-circle of small radius at $ x \geq 0 $,  we obtain
from (\ref{54})-(\ref{56}) :

\begin{equation}\label{58}
 v _ z (z, \:\bar z) + k _ {-} v (z, \:\bar z) = \int _ {\Gamma\bigcup
C _ R} \frac {d\zeta} {2\pi i} \:\frac {v _ {\zeta} (\zeta, \:
{\bar \zeta}) + k _ {-} v (\zeta, \:\bar \zeta)} {\zeta-z} - \int
_ {\mathbb {D}} \int \frac {d\zeta \wedge d {\bar \zeta}} {2\pi i}
\:\frac {k _ {+} v _ {\zeta} (\zeta, \: {\bar \zeta})} {\zeta-z},
\end{equation}

\begin{equation}\label{59}
v_ {\bar z} (z, \:\bar z) + k _ {+} v (z, \:{\bar z)} = -\int _
{\Gamma \bigcup C _ R} \frac {d {\bar \zeta}} {2\pi i} \:\frac {v
_ {\bar \zeta} (\zeta, \: {\bar \zeta}) + k _ {+} v (\zeta, \:\bar
\zeta)} {{\bar\zeta} - {\bar z}} -\int _ {\mathbb {D}} \int \frac
{d \zeta \wedge d {\bar \zeta}} {2\pi i} \:\frac {k _ {-} v _
{\bar \zeta} (\zeta, \: {\bar \zeta})} {{\bar \zeta} - {\bar z}} .
\end{equation}
These relations can be also simplified. Indeed, taking into
account (\ref{51}) and (\ref{57}) we have:

\begin{equation}\label{60}
 \lim _ {R \to \infty} \int _ {C _ R} \frac {d\zeta} {2\pi i}
\:\frac {v _ {\zeta} (\zeta, \: {\bar \zeta}) + k _ {-} v (\zeta,
\: {\bar \zeta})} {\zeta-z} = \lim _ {R \to \infty} \int _ {C _ R}
\frac {d {\bar \zeta}} {2\pi i} \:\frac {v _ {\bar \zeta} (\zeta,
\: {\bar \zeta}) + k _ {+} v (\zeta, \: {\bar \zeta})} {{\bar
\zeta} -\bar z} = 0.
\end{equation}
Besides that, condition (\ref{50}) gives:

\begin{equation}\label{61}
 \int _ {\Gamma} \frac {d\zeta} {2\pi i} \:\frac {v _ {\zeta}
(\zeta, \: {\bar \zeta}) + k _ {-} v (\zeta, \:\bar \zeta)}
{\zeta-z} = \int _ {-\infty}^0\frac {d\xi} {2\pi i} \:\frac {v ^
{+} _ {\zeta} (\xi)-v ^{-}_{\zeta} (\xi)} {\xi-z},
\end{equation}
\begin{equation}\label{62}
 \int _
{\Gamma} \frac {d {\bar \zeta}} {2\pi i} \:\frac {v _ {\bar \zeta}
( \zeta, \: {\bar \zeta}) + k _ {+} (\zeta, \:\bar \zeta)} {{\bar
\zeta} - {\bar z}} = \int _ {-\infty} ^ 0 \frac {d\xi} {2\pi i}
\:\frac {v ^ {+}_ {\bar \zeta} (\xi) -v ^ {-} _ {\bar \zeta}
(\xi)} {\xi - {\bar z}}.
\end{equation}
 Then equations (\ref{58})-(\ref{59}) yield:

\begin{equation}\label{63}
 v _ z (z, \:\bar z) + k _ {-} v (z, \:\bar z)  = \int _
{-\infty} ^ 0 \frac {d\xi} {2\pi i} \frac {v ^ {+}_ {\zeta} (\xi)
-v ^ {-} _ {\zeta} (\xi)} {\xi-z} \: - k _ {+} \int _ {\mathbb D}
\int \:\frac {d\zeta \wedge d {\bar \zeta}} {2\pi i} \frac {v _
{\zeta} (\zeta, \:\bar \zeta)} {\zeta-z},
\end{equation}
\begin{equation}\label{64}
 v _ {\bar z} (z, \:\bar
z) + k _ {+} v (z, \:\bar z) = -\int _ {-\infty} ^ 0 \:\frac
{d\xi} {2\pi i} \frac {v ^ {+} _ {{\bar \zeta}} (\xi) -v ^ {-} _
{\bar \zeta} (\xi)} {\xi - {\bar z}} \: - k _ {-} \int _ {\mathbb
D} \int \:\frac {d \zeta \wedge d {\bar \zeta}} {2\pi i} \frac {v
_ {\bar \zeta} (\zeta, \:\bar \zeta)} {{\bar \zeta} -\bar z} ,
\end{equation}
where we used notations:

\begin{equation}\label{65}
 v ^+ _ {\zeta} (\xi) = \frac {1} {2} (\partial _ {\xi}
-i\partial _ {\eta}) v (\xi + i\eta, \:\xi -i\eta) _ {| \eta \to
\pm 0}, \:\: \:v ^+ _ {\bar \zeta} (\xi) = \frac {1} {2}
(\partial _ {\xi} + i\partial _ {\eta}) v (\xi + i\eta,
\:\xi-i\eta) _ {| \eta \to \pm 0}.
\end{equation}

 The equations (\ref{63})-(\ref{64}) allow, in particular, to obtain
the system of equations which connects a derivative of a field on
a negative half-axis with derivative in the domain $ \mathbb D $.
Really, for $ z = x < 0 $ and taking into account the boundary
condition (\ref{50}), equations (\ref{63}) can be written as

\begin{equation}\label{66}
 v _ z ^ {+}(x)-k _ {-} = \int _ {-\infty} ^ 0 \frac {d\xi _ 1}
{2\pi i} \:\frac {v _ {\zeta _ 1} ^ {+} (\xi _ 1) -v _ {\zeta _ 1}
^ {-} (\xi _ 1)} {\xi _ 1-x -i0} -k _ {+} \int _ {\mathbb D} \int
\frac {{d\zeta_1} \wedge d {\bar \zeta _ 1}} {2\pi i} \:\frac {v _
{\zeta _ 1}(\zeta _ 1, \:\bar \zeta _ 1)} {\zeta _ 1-x},
\end{equation}

\begin{equation}\label{67}
 v _z ^{-}(x)-k_{-} = \int _ {-\infty} ^ 0 \frac {d\xi _ 1}
{2\pi i} \:\frac {v _ {\zeta _ 1} ^ {+} (\xi _ 1) -v _ {\zeta _ 1}
^ {-} (\xi _ 1)} {\xi _ 1-x + i0} -k _ {+} \int _ {\mathbb D} \int
\frac {{d\zeta _ 1} \wedge d {\bar \zeta _ 1}} {2\pi i} \:\frac {v
_ {\zeta _ 1}(\zeta _ 1, \:\bar \zeta _ 1)} {\zeta _ 1-x},
\end{equation}
 and equation (\ref{64}) as

\begin{equation}\label{68}
 v _{\bar z}^{+}(x)-k_{+}=-\int _ {-\infty} ^ 0 \frac {d\xi _ 2} {2\pi i}
\:\frac {v _ {\bar \zeta _ 2} ^ {+} (\xi _ 2) -v _ {\bar \zeta _
2} ^ {-} (\xi _ 2)} {\xi _ 2-x + i0} -k _ {-} \int _ {\mathbb D}
\int \frac {d {\zeta _ 2} \wedge d {\bar { \zeta _ 2}}} {2\pi i}
\:\frac {v _ {{\bar \zeta _ 2}}(\zeta _ 2, \:\bar \zeta _ 2)}
{{\bar \zeta _ 2} -x},
\end{equation}
\begin{equation}\label{69}
 v _{\bar z}^{-}(x)-k_{+}=-\int _ {-\infty} ^ 0 \frac
{d\xi _ 2} {2\pi i} \:\frac {v _ {\bar \zeta _ 2} ^ {-} (\xi _ 2)
-v _ {\bar \zeta _ 2} ^ {-} (\xi _ 2)} {\xi _ 2-x -i0} -k _ {-}
\int _ {\mathbb D} \int \frac {d {\zeta _ 2} \wedge d {\bar {
\zeta _ 2}}} {2\pi i} \:\frac {v _ {{\bar \zeta _ 2}}(\zeta _ 2,
\:\bar \zeta _ 2)} {{\bar \zeta _ 2} -x}.
\end{equation}

The system of equations (\ref{66})-(\ref{69}) is closed system of
singular integral equations on the axis $ x, $ to find the
functions $ v ^{ +}_z,\:v^{-}_z$ and $v^{+}_{\bar z},\:v^{-}_{\bar
z},$ respectively, from the values of $ v_z $ and $ v _ {\bar z} $
in an arbitrary point of domain $ \mathbb D . $

These functions can be also calculated using another approach.
Indeed, equation (\ref{49}) can be rewritten as

\begin{equation}\label{70}
(\frac {1} {2} \:\partial _ {\bar z} + k _ {+}) v _ z (z, \:\bar
z) = - (\frac {1} {2} \:\partial _ z + k _ {-}) v _ {\bar z} (z,
\:\bar z).
\end{equation}
Then we construct the generalized function - fundamental solution
of the operator $ ((1/4) (\partial _ x + i\partial _ y) + k _ {+})
$, which satisfies the equation

\begin{equation}\label{71}
 [ (\partial _ x + i\partial _ y) + 4k _ {+}] G _ 1 (x, \:y, \:k
_ {+}) = 4\delta (x) \delta (y).
\end{equation}

Thus from the equation (\ref{70}) it follows:

\begin{equation}\label{72}
v _ z (x, \:y) = -\int _ {\mathbb D} \int\:d\xi \; d\eta\:G _ 1
(x-\xi, \:y-\eta, \:k _ {+}) [\partial _ {\xi} -i\partial _ {\eta}
+ 4k _ {-}] v _ {\bar \zeta} (\xi, \:\eta).
\end{equation}

Similarly we can find the generalized function - fundamental
solution of the operator $ ((1/4) (\partial _ x-i\partial _ y) + k
_ {-}) $:

\begin{equation}\label{73}
 [ (\partial _ x-i\partial _ y) + 4k _ {-}] G _ 2 (x, \:y, \:k _
{-}) = 4\delta (x) \delta (y).
\end{equation}

From (\ref{70}) and (\ref{73}) we obtain:

\begin{equation}\label{74}
 v _ {\bar z} (x, \:y) = -\int _ {\mathbb D} \int\:d\xi \;
d\eta\:G _ 2 (x-\xi, \:y-\eta, \:k _ {-}) [ \partial _ {\xi} +
i\partial _ {\eta} + 4k _ {+} \:] v _ {\zeta} (\xi, \:\eta).
\end{equation}

The solutions $G_1$ and $G_2 $ are connected among themselves by
the symmetry relation:

\begin{equation}\label{75}
 G _ 1 (z, \:\bar z, \:k _ {+}) = - {\bar G}_2 (-z, \:\bar z,
\: - {\bar k} _ {+}), \:\:\:\: G _ 2 (z, \:\bar z, \:-k _ {-}) = -
{\bar G} _ 1 (-z, \: - {\bar z}, \:-{\bar k} _ {-}).
\end{equation}

The system of equations (\ref{72}), (\ref{74}) is a system of
integral equations of the convolution type, and it can be solved
by the Fourier transformation (for generalized functions). Such
calculations as well as the explicit expressions for functions
$G_1,\: G_2$ and the comparison with the results of \cite{14} will
be given later.

The approach developed above in this Section is seems to be an
alternative to the traditional Wiener-Hopf approach. It would be
interesting to generalize this method for the case of
three-dimensional problem of diffraction.

\vskip0.3cm

{\bf {Acknowledgements}} \vskip0.3cm \vskip0.3cm

The author is grateful to M.V.Ioffe for attention and A.V.Tukhtin
for useful discussions of questions of the theory of diffraction.
 
\end {document}